\documentclass[10pt,twocolumn,letterpaper]{article}

\usepackage[pagenumbers]{cvpr} %

\usepackage[dvipsnames]{xcolor}

\makeatletter
\def\adl@drawiv#1#2#3{%
        \hskip.5\tabcolsep
        \xleaders#3{#2.5\@tempdimb #1{1}#2.5\@tempdimb}%
                #2\z@ plus1fil minus1fil\relax
        \hskip.5\tabcolsep}
\newcommand{\cdashlinelr}[1]{%
  \noalign{\vskip\aboverulesep
           \global\let\@dashdrawstore\adl@draw
           \global\let\adl@draw\adl@drawiv}
  \cdashline{#1}
  \noalign{\global\let\adl@draw\@dashdrawstore
           \vskip\belowrulesep}}
\makeatother

\usepackage{times}
\usepackage{epsfig}
\usepackage{graphicx}
\usepackage{amsmath}
\usepackage{amssymb}
\usepackage{booktabs}
\usepackage{multirow}
\usepackage{xcolor}
\usepackage{arydshln}
\usepackage{wrapfig}
\usepackage{enumitem}
\usepackage[font=small]{caption}
\usepackage{tabularx} 
\usepackage{microtype}
\usepackage{pifont}
\usepackage{listings}

\definecolor{codegreen}{rgb}{0,0.6,0}
\definecolor{codegray}{rgb}{0.5,0.5,0.5}
\definecolor{codepurple}{rgb}{0.58,0,0.82}
\definecolor{backcolour}{rgb}{0.95,0.95,0.92}

\lstdefinestyle{mystyle}{
    backgroundcolor=\color{backcolour},   
    commentstyle=\color{codegreen},
    keywordstyle=\color{magenta},
    numberstyle=\tiny\color{codegray},
    stringstyle=\color{codepurple},
    basicstyle=\ttfamily\footnotesize,
    breakatwhitespace=false,         
    breaklines=true,                 
    captionpos=b,                    
    keepspaces=true,                 
    numbers=left,                    
    numbersep=5pt,                  
    showspaces=false,                
    showstringspaces=false,
    showtabs=false,                  
    tabsize=2,
    showlines=true
}

\newcommand{\xmark}{\ding{55}}
\newcommand{\cmark}{\ding{51}}

\newcommand\rgt{\aftergroup\mathclose\aftergroup{\aftergroup}\right}

\newcommand{\ourdata}{RAF\xspace}

\newcommand{\supparxiv}[2]{#2}
\def\upvspacefig{\vspace{-0mm}}
\def\downvspacefig{\vspace{-0mm}}
\makeatletter
\renewcommand\paragraph{\@startsection{paragraph}{4}{\z@}%
	{0.75ex \@plus.5ex \@minus.2ex}%
	{-1em}%
	{\normalfont\normalsize\bfseries\maybe@addperiod}}
\newcommand{\maybe@addperiod}[1]{#1\@addpunct{.}}
\makeatother

\def\projecturl{https://facebookresearch.github.io/real-acoustic-fields}

\definecolor{cvprblue}{rgb}{0.21,0.49,0.74}
\definecolor{urlblue}{rgb}{0.24,0.49,0.9}
\usepackage[pagebackref,breaklinks,colorlinks,urlcolor=urlblue,citecolor=cvprblue]{hyperref}

\def\paperID{1631} %
\def\confName{CVPR}
\def\confYear{2024}

\title{Real Acoustic Fields: An Audio-Visual Room Acoustics Dataset and Benchmark}

\author{Ziyang Chen\textsuperscript{1,2$^{*}$}\qquad 
Israel D. Gebru\textsuperscript{2}\qquad 
Christian Richardt\textsuperscript{2} \qquad
Anurag Kumar\textsuperscript{3} \vspace{1.2mm}\\
William Laney\textsuperscript{2} \qquad 
Andrew Owens\textsuperscript{1} \qquad 
Alexander Richard\textsuperscript{2}
\vspace{2.5mm}\\
\textsuperscript{1}University of Michigan\quad \textsuperscript{2}Codec Avatars Lab, Pittsburgh, Meta \quad
\textsuperscript{3}Reality Labs Research, Meta
\vspace{1.25mm}\\
\normalsize{\url{\projecturl}}
}

\begin{document}
\maketitle

{\let\thefootnote\relax\footnotetext{{* Work done during an internship at Meta.}}}

\begin{abstract}
We present a new dataset called Real Acoustic Fields (\ourdata) that captures real acoustic room data from multiple modalities. The dataset includes high-quality and densely captured room impulse response data paired with multi-view images, and precise 6DoF pose tracking data for sound emitters and listeners in the rooms. We used this dataset to evaluate existing methods for novel-view acoustic synthesis and impulse response generation which previously relied on synthetic data. In our evaluation, we thoroughly assessed existing audio and audio-visual models against multiple criteria and proposed settings to enhance their performance on real-world data. We also conducted experiments to investigate the impact of incorporating visual data (i.e., images and depth) into neural acoustic field models. Additionally, we demonstrated the effectiveness of a simple sim2real approach, where a model is pre-trained with simulated data and fine-tuned with sparse real-world data, resulting in significant improvements in the few-shot learning approach. \ourdata is the first dataset to provide densely captured room acoustic data, making it an ideal resource for researchers working on audio and audio-visual neural acoustic field modeling techniques. Demos and datasets are available on our \href{\projecturl}{project page}.

\end{abstract}

\supparxiv{\vspace{-2.0mm}}{}
\section{Introduction}
\label{sec:intro}

\begin{figure*}[!t]
    \centering
    \upvspacefig
    \supparxiv{\vspace{-5.0mm}}{}
    \includegraphics[width=\linewidth]{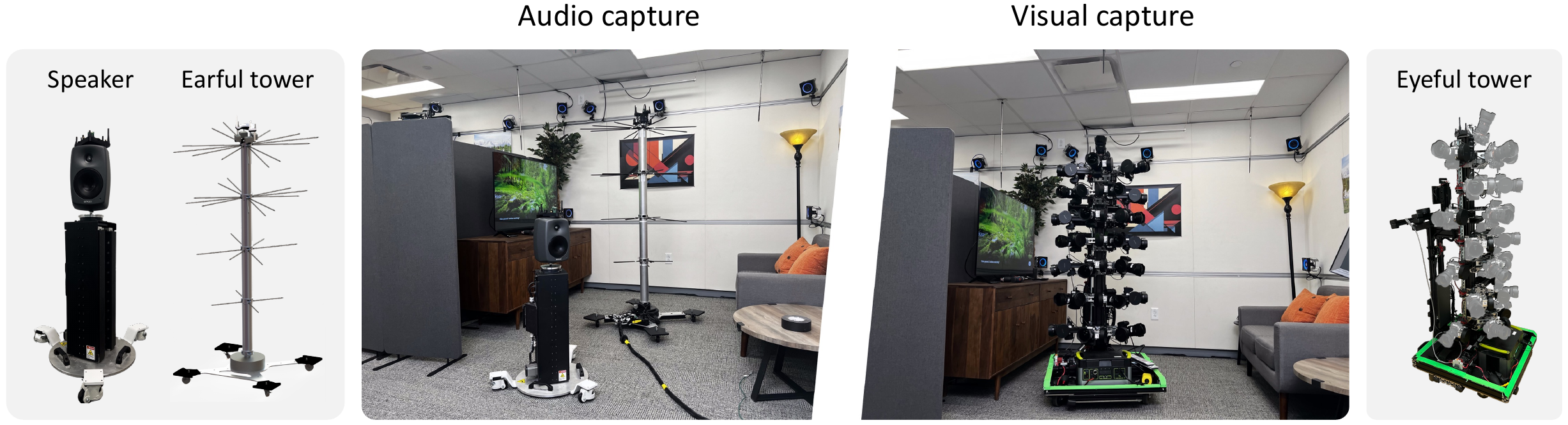}
    
    \caption{\textbf{Data capturing setup.}
        (a) Audio capture (left): the loudspeaker and microphone recording system~(Earful Tower) are placed at different locations within the room to measure and capture RIRs.
        (b) Visual capture (right): the camera rig~(Eyeful Tower) moves around rooms to capture multi-view images for visual reconstruction and novel-view synthesis.
    }
    \label{fig:data_overview}
    \downvspacefig
\end{figure*}

Sound waves reflect off objects in a scene before reaching a listener's ears. These reflections change the sound waves in complex ways and convey the objects' size, shape, and material properties. Accurately modeling these changes is crucial for spatial audio rendering, and plays a key role in adding the sense of immersion in a variety of application domains, such as 3D games, virtual and augmented reality~\cite{zhang2017surround, hammershoi2005binaural}.

The goal of a sound propagation model is typically to estimate a room impulse response (RIR) for a given emitter and listener pose. RIRs are linear filters that, when convolved with an input sound, simulate the sound that would be perceived by the listener, performing changes like adding reverb or dampening certain frequencies. Estimating RIRs for novel emitter and listener poses from sparsely sampled RIRs acquired from a scene has been a major focus of recent work in audio~\cite{nakamura2000acoustical,jeub2009binaural,hadad2014multichannel,ko2017study,carlo2021dechorate} and audio-visual learning~\cite{chen2022visual,singh2021image2reverb,somayazulu2023self, majumder2022few,chen2023everywhere, AhnYHSRSTC2023}. 
Inspired by novel-view synthesis~\cite{mildenhall2021nerf, martin2021nerf, li2022neural}, an emerging line of work has proposed learning-based models based on neural fields~\cite{luo2022learning, liang2023neural, SuCS2022}. 

Despite the recent interest in sound propagation in the audio-visual community, existing methods have been developed and evaluated on highly simplified datasets with artificially generated impulse responses. This is due to the fact that collecting real-world RIRs is a challenging process that requires both playing and recording sounds from densely sampled positions throughout a scene. Many different data collection efforts have each made different compromises between the conflicting factors in terms of realism, ground truth, and costs.
Existing datasets~\cite{chen2023novel,liang2023av,koyama2021meshrir} thus make highly restrictive assumptions, such as by having only a single sound emitter at a fixed pose, by having limited (2D-only) spatial coverage of the scenes, or by having only simple planar geometry.
Consequently, these datasets do not fully capture the complexities of real-world room geometry, material variations, and source directivity. %
The lack of a real ``gold standard'' benchmark makes it challenging to effectively analyze existing approaches under real-world assumptions and to drive research on audio-visual informed sound propagation toward its true potential.

In this paper, we propose an audio-visual sound propagation dataset and benchmark that addresses the shortcomings of previous approaches. 
Our \emph{Real Acoustic Fields~(\ourdata)} dataset is a %
multimodal real acoustic room dataset with dense 3D audio captures of a large space filled with and without furniture. To capture dense and calibrated audio in the rooms, we used a custom-built microphone tower system and robotic loudspeaker stand. The microphone tower contains 36 omnidirectional microphones placed at different heights and positions. The robotic stand can rotate and position the loudspeaker at different heights, enabling us to capture sound source directivity data. We used a motion capture system to precisely track the pose of the microphones and the loudspeaker throughout the scene. Moreover, we pair our RIR recordings with captured high-fidelity images and geometry~\cite{XuALGBKBPKBLZR2023} to enable more potential research in the audio-visual direction. The resulting dataset contains high-fidelity dense RIRs, speech recordings from existing speech datasets, position annotations, and visual reconstructions.

Using this dataset, we conduct the first systematic study of recent audio and audio-visual sound propagation models, including: 1) extension of common 2D approaches into 3D scenes, 2) how perceptual similarity metrics proposed by other models \cite{majumder2022few,liang2023neural} change the generated sounds, and 3) the role of visual information in audio-visual models.
Finally, our dataset also allows us to evaluate few-shot training. We propose a simple, yet highly effective ``sim2real'' approach that begins by pretraining on synthetic data and then refining the result with a small number of real-world samples. We will release the dataset and benchmark upon acceptance.

\begin{table*}[!t]
\upvspacefig
\supparxiv{\vspace{-2.0mm}}{}
\caption{\textbf{Dataset comparison.} We compare the attributes of our dataset with previously proposed datasets.  }
\label{tab:data-comparison}

\centering
\small
\begin{tabular}{lccccccc}

\toprule
Dataset & Modality &  Real-world & Visual source & Dimension & Scenes & Density & \\
\midrule
SoundSpaces 1.0~\cite{ChenJSGAIRG2020} & $\mathcal{A}$ \& $\mathcal{V}$  & \xmark &   Mesh &  2D & 103 & 16 samples/m$^2$\\
SoundSpaces 2.0~\cite{chen2022soundspaces} & $\mathcal{A}$ \& $\mathcal{V}$  & \xmark & Mesh  &  3D & 1600+ & --\\

MeshRIR~\cite{koyama2021meshrir} & $\mathcal{A}$ & \cmark &  -- & 2.5D & 1 &  18 samples/m$^3$\\
\ourdata~(ours)& $\mathcal{A}$ \& $\mathcal{V}$  & \cmark &   NeRF \& Mesh & 3D & 2 &  372 samples/m$^3$\\
\bottomrule
\end{tabular}
\supparxiv{\vspace{-3.5mm}}{}
\end{table*}

\section{Related Work}
\label{sec:related_work}

\paragraph{Novel-view acoustic datasets}
Many RIR datasets are collected for acoustic research~\cite{jeub2009binaural, le2014categorization, le2015micbots, wang2023soundcam} while they are not applicable for novel-view acoustic propagation modeling. 
MeshRIR~\cite{koyama2021meshrir} recorded real-world monaural impulse responses from a three-dimensional cuboidal room, with microphones at a fixed height. The room was empty and lacked visual information about the scene.
Two previous methods by \citet{liang2023av} and \citet{chen2023novel} collected real-world audio-visual datasets for novel-view acoustic synthesis tasks.
Nevertheless, \citet{liang2023av} only features a single stationary sound source and \citet{chen2023novel} only has sparse receiver positions, which might not represent the entire acoustic environment for arbitrary speaker-receiver pairs.
SoundSpaces 1.0~\cite{ChenJSGAIRG2020} and SoundSpaces 2.0~\cite{chen2022soundspaces} generated large-scale synthetic acoustic datasets based on room mesh from existing 3D scene datasets~\cite{ChangDFHNSSZZ2017,straub2019replica,xia2018gibson,RamakGWMCTUGWCSZB2021,yadav2023habitat}.
However, these synthetic datasets lack the complexities of real-world room geometry, material variations, and source directivity.
To address this, we have gathered a real-world multimodal acoustic room dataset to further research in the field of neural acoustics.

\paragraph{RIR synthesis}
Synthesizing RIR has been a longstanding research topic.
Simulated approaches for RIR synthesis primarily rely on wave-based~\cite{thompson2006review, gumerov2009broadband} or geometric methods~\cite{schissler2016interactive, chen2020soundspaces, chen2022soundspaces}.
While these methods effectively simulate sound propagation in space, they often struggle to reproduce all wave-based sound effects. Geometric models do not account for interference and diffraction. While wave-based models are theoretically applicable to all frequencies, they face difficulty in accurately modeling the frequency-dependent directional characteristics of sound sources, receivers, and rooms with complex geometries.
Recent methods have leveraged machine learning techniques to create more realistic RIRs.
\citet{ratnarajah2022fast} use a generative adversarial network~(GAN) to synthesize RIRs.
Later work extended this approach by conditioning on scene meshes~\cite{ratnarajah2022mesh2ir} and visual signals~\cite{ratnarajah2023av}.
Few other works focus on learning continuous implicit neural representations for audio scenes, which target generating high-fidelity impulse responses at any arbitrary emitter-listener positions for a single scene, such as NAF~\cite{luo2022learning}, INRAS~\cite{SuCS2022} and NACF~\cite{liang2023neural}.
Nevertheless, prior studies have primarily focused on simulated data owing to the absence of suitable real-world datasets.
Our novel dataset presents a path to extend these approaches toward real-world modeling of neural acoustic fields.

\paragraph{Audio-visual acoustic learning} 

Recent works have explored learning acoustic information from both audio and vision.
\citet{chen2023learning} and \citet{chowdhury2023adverb} propose de-reverberating audio signals using visual environment encoding.
Some researchers investigate the visual acoustic matching problem \cite{chen2022visual,singh2021image2reverb,somayazulu2023self}, aiming to synthesize audio that matches target acoustic properties based on images.
Some works learn to generate sounds at the arbitrary speaker and listener positions via sparse audio-visual observations of scenes~\cite{majumder2022few,chen2023everywhere}. Others focus on the novel-view acoustic synthesis task, synthesizing binaural sound from audio and visual information at a new viewpoint~\cite{chen2023novel,liang2023av, AhnYHSRSTC2023,chen2023sound}.
We use our dense 3D audio-visual dataset to evaluate these methods' effectiveness and the role of vision.

\paragraph{Visual scene capture and view synthesis}
There is a rich literature on capturing static scenes to reconstruct them in 3D and/or to render novel viewpoints; see recent surveys for a comprehensive overview~\cite{TewarTMSTWLSMLSTNBWZG2022, RichaTW2020}.
Many approaches that focus on 3D scene reconstruction use representations such as (truncated) signed distance fields to combine multiple observations from RGB-D sensors ~\cite{NewcoDIKHSMHKF2011, NiessnZIS2013, YangLCFLKH2020, SucarLOD2021}
or standard color videos~\cite{MurezABSBR2020, SunXCZB2021, GuoPLWZBZ2022, JangMKKRK2022}.
These approaches tend to sacrifice rendering fidelity in favor of better 3D reconstruction accuracy.
On the other hand, when the visual quality of novel views is paramount, approaches building on image-based rendering~\cite{ShumCK2007, HedmaRDB2016, OverbEEPD2018, BerteYLR2020}
or, more recently, neural radiance fields \cite{MildeSTBRN2020, WuXZBHTX2022, BarroMVSH2023, XuALGBKBPKBLZR2023} have achieved the highest visual fidelity, even while compromising the quality of the reconstructed 3D geometry.
To maximize the visual fidelity of our dataset, we capture and reconstruct it using the VR-NeRF approach~\cite{XuALGBKBPKBLZR2023}.

\section{The \ourdata Dataset}
\label{sec:dataset}

We present \ourdata, a dataset of densely recorded real-world room impulse responses~(RIR) paired with dense multi-view images of the scenes.
To the best of our knowledge, this is the first multi-modal 3D RIR dataset with dense audio and visual measurements paired with precise 6DoF tracking data.
In this section, we will introduce the hardware setup used for data collection and our data collection pipeline.

\subsection{Audio Capturing} 

Our goal is to collect dense RIR samples that cover the entire scene with paired transmitter and receiver locations.

\paragraph{Hardware} 
To facilitate the audio data collection process, we developed a novel microphone tower system called \emph{Earful Tower}, as shown in \cref{fig:data_overview}.
The tower features 36 omnidirectional microphones.
These microphones were placed at different height levels on the tower, arranged in the shape of an inverted pine cone.
We positioned more microphones at the average human ear height level and used fewer microphones at lower levels.
The microphones are integrated with three RME 12Mic-D units, daisy-chained and phase-locked to record synchronized multi-channel audio signals.

For generating room excitation signals during RIR measurements, we used a Genelec 8030C speaker mounted on a robotic stand.
This stand offers remote control, programmable height adjustment, and speaker axis rotation.

\paragraph{Capturing procedure}
We uniformly distribute the microphone tower at walkable positions in the room that might be occupied by a human listener~(\ie, open areas). We used the robotic stand to automate the rotation of the loudspeaker every 120° on its axis at each position, to obtain differently oriented sound sources. During the recording process, we played logarithmic sine-sweep signals and simultaneously recorded the resulting reverberated signals using the microphones on the tower. After completing a full-circle rotation, the speaker stand would adjust its height, and we would repeat the measurements for each 120° turn. Then we relocated the microphone tower to a new position and repeated the measurements. We shifted the speaker to a new location after the microphone tower had swept through the entire scene. Meanwhile, after each sine-sweep, we played and recorded 6-second long speech utterances randomly sampled from the VCTK dataset \cite{veaux2013voice}.%

To accurately track the orientation and positions of the loudspeaker and microphones in the room, we used the OptiTrack motion capture system.
We placed reflective markers on the loudspeaker and the microphone tower, allowing us to precisely estimate their 6 degrees of freedom (6DoF) poses.

\begin{figure}[!t]
    \centering
    \upvspacefig
    \supparxiv{\vspace{4.0mm}}{}
    \includegraphics[width=\linewidth]{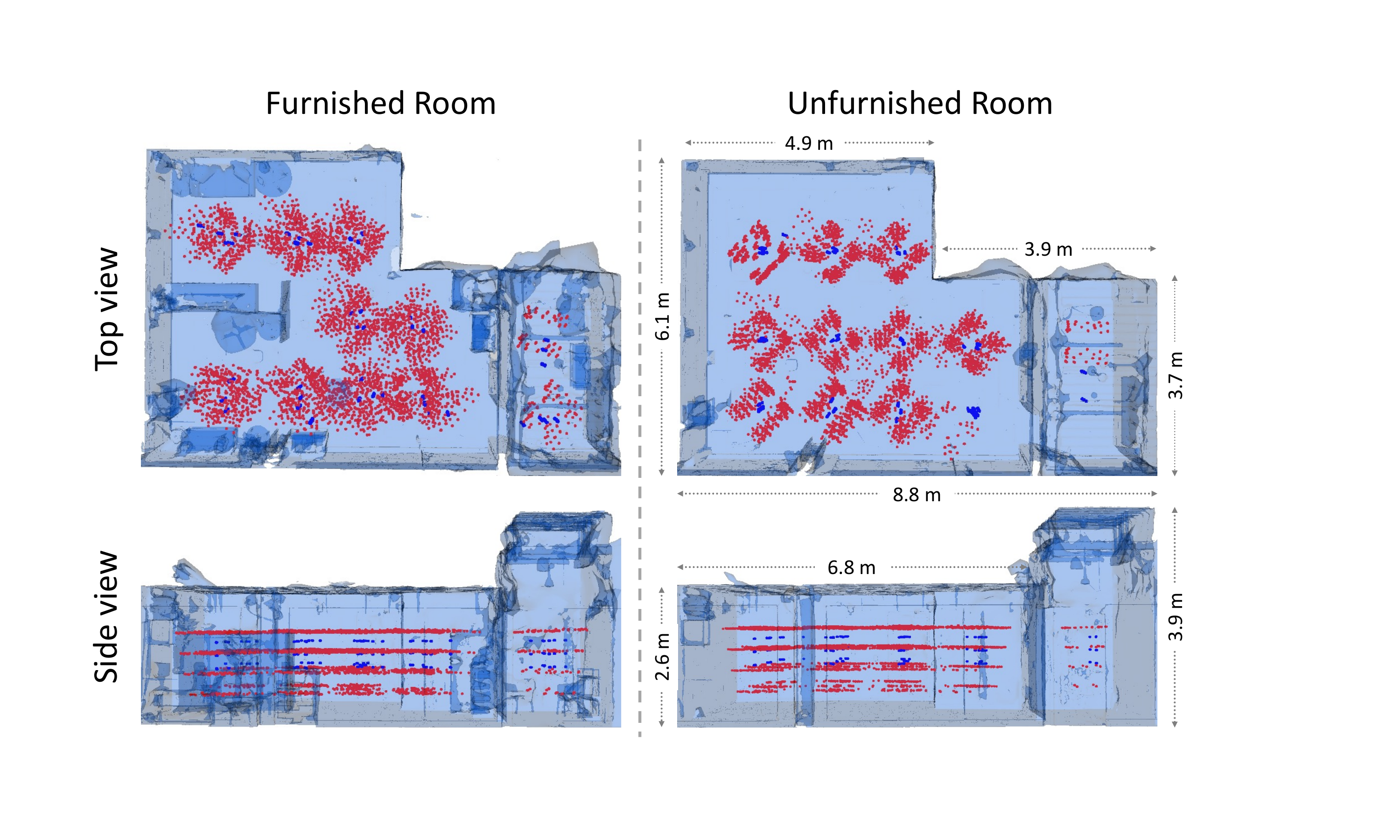}
    \caption{\textbf{Data distribution of \ourdata.} Blue dots represent speaker positions and red dots represent microphone positions. The room dimensions are shown on the right.} 
    \label{fig:data_distribution}
    \downvspacefig
    \vspace{-2mm}
\end{figure}

\paragraph{Captured data}
With the setup described above, we collected dense data from one room under two different configurations. In the first setting, the room was empty and only contained the essential equipment necessary for capturing impulse response data. In the second setting, we furnished the room to resemble a simple studio or living room. We collected 47K RIRs for the empty room and 39K RIRs for the furnished room. The collected RIRs are \textbf{4 seconds long}, which comprehensively captures the acoustic information. Our room has two parts: a large room with soft material walls to absorb sounds, and a smaller room with concrete walls for increased reverberation.
We show our RIR distribution and room measurements in \cref{fig:data_distribution}. 
Please see \supparxiv{the supplement}{\cref{supp:dataset}} for more details.

\subsection{Visual Capturing}
To provide a high-fidelity visual reconstruction of the scenes and synthesize the appearance from any viewpoint, we follow the VR-NeRF approach \cite{XuALGBKBPKBLZR2023} to capture dense multi-view images for the scenes, using the \emph{Eyeful Tower} multi-camera rig shown in \cref{fig:data_overview}.
We move the Eyeful Tower rig to cover the available floor area for a dense capture of our static scenes, resulting in 3,388 images for the furnished room and 8,030 images for the empty room.
We use Agisoft Metashape \cite{Agiso2023} to estimate the poses of cameras within the rig using structure-from-motion and reconstruct a textured mesh of each scene.
Lastly, we use ground control points to align the cameras and RIR data to the same coordinate system.
The audio and visual captures are performed separately to prevent any interference between audio and visual devices (\eg, speakers and microphones appearing in the images) and to eliminate the impact of camera devices on audio capture (\eg, cameras creating reflections).

To generate the views at each microphone or speaker position, we train NeRF models using the Instant NGP architecture \cite{XuALGBKBPKBLZR2023} for each scene. This enables us to examine the effectiveness of incorporating visual signals, such as RGB and depth information, into acoustic field modeling.

\subsection{Comparison to Prior Datasets} 

We compare our dataset to several prior acoustic datasets collected from real scenes or through simulators in \cref{tab:data-comparison}.
In comparison to the real MeshRIR dataset~\cite{koyama2021meshrir}, our dataset offers 20 times denser and more extensive coverage of room impulse response data from different height levels.
Furthermore, our dataset features more complex room geometry and materials with furniture, going beyond the limitations of a single box-shaped room.
Compared to simulated datasets such as SoundSpaces 1.0~\cite{chen2020soundspaces} and SoundSpaces 2.0~\cite{chen2022soundspaces}, 
our dataset stands out for its high-quality real impulse responses and high-fidelity visual rendering from NeRF. In contrast, the simulated datasets fall behind in terms of both audio and visual quality, resulting in a less realistic representation of real-world acoustics.

\section{Learning 3D Neural Acoustic Fields}

Modeling acoustic fields can be formulated as: given the speaker's spatial position $\mathbf{s} = (x_\text{s}, y_\text{s}, z_\text{s}) \!\in\! \mathbb{R}^3$,
the speaker orientation $\boldsymbol\theta \!\in\! \mathbb{R}^2$, and the receiver spatial position $\mathbf{r} = (x_\text{r}, y_\text{r}, z_\text{r}) \in \mathbb{R}^3$ in a room, a function $\mathcal{F}$ predicts the corresponding impulse response $h$:
\begin{equation}
    \mathcal{F}: (\mathbf{s},\mathbf{r}, \boldsymbol\theta ) \mapsto h \in \mathbb{R}^T \text{.}
\end{equation}
Previous studies have proposed various methods to learn $\mathcal{F}$, but they rely on synthetic data. We investigate and improve those existing models in real-world scenarios using our dataset. Additionally, we evaluate the effectiveness of using geometry or visual cues to model acoustic fields. Moreover, we introduce a simple yet effective sim2real approach for synthesizing RIRs in few-shot scenarios, which can significantly enhance performance.

\subsection{Models}
We adopted several existing state-of-the-art 2D acoustic field and audio-visual models to our 3D setup, with some modifications. These models are briefly described below.

\paragraph{NAF}

The neural acoustic field~\cite{luo2022learning} models the room acoustics using an implicit representation.
NAF learns a grid of local geometric features $\mathbf{G}$ to encode the spatial information of speakers and receivers at different positions, and queried speakers and receivers grid features $\mathbf{G(\mathbf{s})}$ and $\mathbf{G(\mathbf{r})}$ will be provided to the NAF $\mathcal{F}$ as additional context.

NAF represents impulse response $h$ in the time-frequency domain $H = | \text{STFT}(h) | \in \mathbb{R}^{F \times K} $ using short-time Fourier transform~(STFT), where $F$ is the numbers of frequency bin and $K$ is the number of time frames. 
Given the frequency bin $f$ and time frame $k$, NAF predicts the log magnitude of the spectrogram $\hat{H}(k, f)$:
\begin{equation}
    \hat{H}(k,f) = \mathcal{F}\left(\mathbf{G(\mathbf{s})}, \mathbf{G(\mathbf{r})}, \boldsymbol\theta, k, f \right) \text{,}
\end{equation}
and it minimizes the $L1$ loss between predicted and ground-truth impulse response in log scale:
\begin{equation}
     \mathcal{L}_\text{NAF} = \Vert \log\hat{H}(k,f) - \log H(k,f) \Vert_1 \text{.}
  \label{eq:naf_loss}
\end{equation}
To obtain the time-domain impulse response $h$, NAF performs inverse STFT on predicted spectrogram magnitude $|\hat{H}|$ with random phase.

\paragraph{INRAS}

The implicit neural representation for audio scenes \cite{SuCS2022} is inspired by interactive acoustic radiance transfer, where sound energy first scatters from the emitter to the boundaries of the scene, then propagates through the scene by bouncing between the surfaces, and finally gathers at the listener position.
INRAS defines a set of bounce points $\{\mathbf{b}_i\}^N_{i=1} \!\subset\! \mathbb{R}^3$, which are uniformly sampled from the scene surface.
It represents the position of speakers or receivers using the relative distance to those bounce points, which provides more information about the scene geometry:
\begin{equation}
    \{\mathbf{d}^{\mathbf{s}}_i\}^N_{i=1} = \{\mathbf{s} - \mathbf{b}_i\}^N_{i=1} \text{,}  \quad \{\mathbf{d}^{\mathbf{r}}_i\}^N_{i=1} = \{\mathbf{r} - \mathbf{b}_i\}^N_{i=1} \text{.}
\end{equation}
INRAS encodes speaker relative distance $\{\mathbf{d}^{\mathbf{s}}_i\}^N_{i=1}$, receiver relative distance $\{\mathbf{d}^{\mathbf{r}}_i\}^N_{i=1}$, and bounce point positions $\{\mathbf{b}_i\}^N_{i=1}$ into latent features $\mathbf{S}, \mathbf{R}, \mathbf{B} \in \mathbb{R}^{N \times D}$, where $D$ is the feature dimension size.
The time embedding $\mathbf{M} \!\in\! \mathbb{R}^{T \times D}$ is introduced for the whole time sequence and obtains spatial-time features via matrix multiplication. 
INRAS generates time-domain RIRs directly by decoding given features:
\begin{equation}
    \hat{h} = \mathcal{F}\left(
        \mathbf{M}\mathbf{S}^\top,
        \mathbf{M}\mathbf{R}^\top,
        \mathbf{M}\mathbf{B}^\top,
        \boldsymbol\theta
    \right) \text{.}
\end{equation}

The INRAS model minimizes the STFT loss in the time-frequency domain $H = |\text{STFT}(h)|$, including spectral convergence loss $\mathcal{L}_\text{sc}$ and magnitude loss $\mathcal{L}_\text{mag}$:
\begin{equation}
\label{eq:inras}
    \mathcal{L}_\text{INRAS}  = \mathcal{L}_\text{sc} + \mathcal{L}_\text{mag} = \frac{\Vert \hat{H} - H \Vert_2}{\Vert  H \Vert_2} + \Vert \hat{H} - H \Vert_1 \text{.}
\end{equation}
We used multi-resolution STFT loss~\cite{yamamoto2020parallel}, which involves computing an STFT loss at multiple time-frequency scales.

\paragraph{NACF}

Neural Acoustic Context Field~(NACF) \cite{liang2023neural} is a multimodal extension of INRAS which uses additional \emph{context} from other modalities.
Specifically, NACF uses RGB images $v_\text{rgb}$ and depth images $v_\text{depth}$ for each predefined bounce point $\mathbf{b}_i$ to extract local geometric and semantic information, and material properties.
Similar to INRAS, RGB and depth images are encoded into latent {\it context} embeddings $\mathbf{C}_\text{rgb}, \mathbf{C}_\text{depth} \in \mathbb{R}^{N \times D}$ via nonlinear projection and converted into space-time features.\footnote{We remove the acoustic coefficient context due to the unavailability of material coefficient annotations for our dataset and our objective of modeling real-world captures without additional annotations.}
NACF decodes provided features with additional context to generate the impulse response in the time domain:
\begin{equation}
\hat{h} = \mathcal{F}\left(
    \mathbf{M}\mathbf{S}^\top,
    \mathbf{M}\mathbf{R}^\top,
    \mathbf{M}\mathbf{B}^\top,
    \mathbf{M}\mathbf{C}_\text{rgb}^\top,
    \mathbf{M}\mathbf{C}_\text{depth}^\top,
    \boldsymbol\theta
\right) \text{.}
\end{equation}

NACF minimizes the same loss as INRAS~(\cref{eq:inras}), as well as the {\em energy decay loss} proposed by \citet{majumder2022few}, which encourages the energy decay curves of the predicted and target RIRs to be similar.
Given the magnitude spectrogram $H \in \mathbb{R}^{F\times K}$, we calculate the decay curve $\mathcal{D}(H)$:
\begin{equation}
    \mathcal{D} (H)[ k ] = 1 + \frac{E_k}{\sum_{i=k+1}^KE_i} \text{,}
\end{equation}
where $E_k = \sum_{f} H(f, k)^2$ is the energy of time frame $k$.
We minimize the $L1$ distance between the predicted and ground-truth decay curves in log space:
\begin{equation}
     \mathcal{L}_\text{decay} = \Vert \log\mathcal{D}(\hat{H}) - \log \mathcal{D}(H) \Vert_1 \text{,}
  \label{eq:decay_loss}
\end{equation}
resulting in the overall loss with multi-resolution STFT:
\begin{equation}
     \mathcal{L}_\text{NACF} =\mathcal{L}_\text{sc} + \mathcal{L}_\text{mag} + \lambda \mathcal{L}_\text{decay} \text{,}
  \label{eq:nacf_loss}
\end{equation}
where $\lambda$ is the weight of the decay loss.

\begin{figure}[!t]
    \centering
    \upvspacefig
    \supparxiv{\vspace{4.0mm}}{}
    \includegraphics[width=\linewidth]{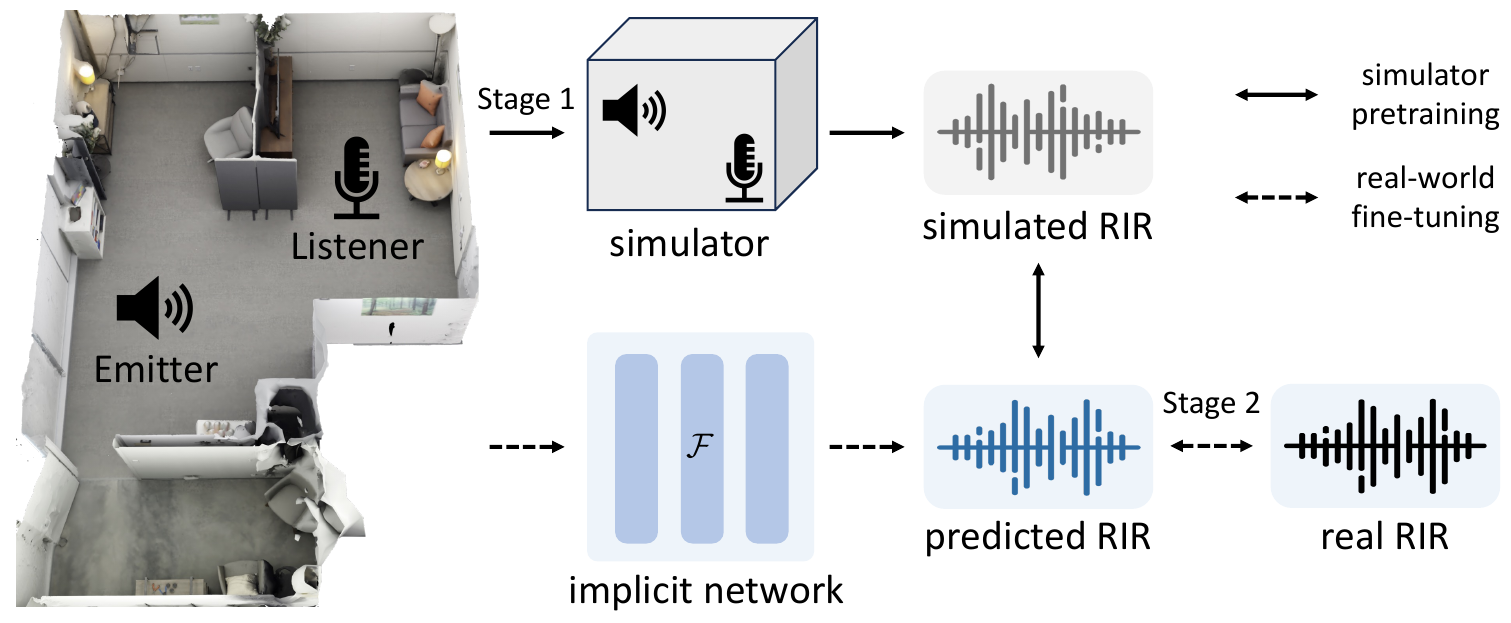}
    
    \caption{\textbf{Sim2real method overview.}
    First, we train the implicit network on simulated data with densely sampling emitter--listener position pairs.
    We then fine-tune it on sparse real-world data.
    } 
    \label{fig:sim2real_method}
\end{figure}

\paragraph{AV-NeRF}

We also consider the very recent AV-NeRF model~\cite{liang2023av}.
Unlike NACF, which uses fixed visual contexts independent from speaker-receiver positions, AV-NeRF provides local visual information $\mathbf{C}_\text{rgb}^{\mathbf{r}}, \mathbf{C}_\text{depth}^{\mathbf{r}}$ that depends on the listener position.
It predicts impulse responses:
\begin{equation}
    \hat{h} = \mathcal{F}\left(\mathbf{s}, \mathbf{r},  \mathbf{C}_\text{rgb}^{\mathbf{r}}, \mathbf{C}_\text{depth}^{\mathbf{r}}, \boldsymbol\theta \right) \text{.}
\end{equation}
We also minimize the losses in \cref{eq:nacf_loss}.

\paragraph{NAF$++$ and INRAS$++$}

We observed that the energy decay loss~(\cref{eq:decay_loss}) from \citet{majumder2022few} used in NACF can also improve the results of the other models.
We therefore introduce improved models NAF$++$ and INRAS$++$ that have these losses:
\begin{align}
	\mathcal{L}_\text{NAF++} &= \mathcal{L}_\text{mag} + \lambda \mathcal{L}_\text{decay} \text{,}\\
    \mathcal{L}_\text{INRAS++} &= \mathcal{L}_\text{sc} + \mathcal{L}_\text{mag} + \lambda \mathcal{L}_\text{decay} \text{.}
\end{align}

\begin{table*}[!t]
\upvspacefig
\supparxiv{\vspace{-1.0mm}}{}
\caption{{\bf Evaluation on \ourdata with 48\,kHz high-fidelity impulse responses.} We evaluate each method with the quality of generated impulse response, storage requirements, and inference speed. The results are averaged across two scenes. Original denotes uncompressed audio. The best results are in {\bf bold}. }
\label{tab:benchmark_48k}

\centering
\footnotesize
\renewcommand{\arraystretch}{1.0}
\newcommand\padd{\phantom{0}}

\resizebox{0.99\textwidth}{!}{
\begin{tabular}{cll@{}cccccrr}

\toprule
 & \multirow{2}{*}{Method} & \multirow{2}{*}{Variation} &  STFT error & $C_{50}$ error & EDT error & $T_{60}$ error & Parameters &  Storage & Speed \\
 & & & (dB)~$\downarrow$ & (dB)~$\downarrow$ & (sec)~$\downarrow$ & (\%)~$\downarrow$ & (Million)~$\downarrow$ & (MB)~$\downarrow$ & (ms)~$\downarrow$ \\
\midrule

\parbox[t]{2mm}{\multirow{6}{*}{\rotatebox[origin=c]{90}{\shortstack[c]{Classical}}}} 
&  \multirow{3}{*}{Linear} & AAC & 1.26   & 2.49  & 0.085  &  25.64 & \multirow{3}{*}{--} & 2,033.81 & \multirow{3}{*}{--} \\
& & Opus &  0.92 & 0.86 & 0.029 & 10.19 &  &2,033.81  &  \\
& & original &  0.88  & 0.83   & 0.027  & \padd 7.82 &  &  9,518.32  &  \\
\cdashlinelr{2-10}
&  \multirow{3}{*}{Nearest} & AAC &  1.04  &  1.97 & 0.064 & 22.83  & \multirow{3}{*}{--} & 2,033.81 & \multirow{3}{*}{--} \\
& & Opus & 0.49 &  0.76 & 0.021  & 10.03 &  & 2,033.81 &  \\
& & original &  0.38  & 0.71  & 0.020  & \padd 7.67 &  &  9,518.32  &  \\

\midrule
\parbox[t]{2mm}{\multirow{8}{*}{\rotatebox[origin=c]{90}{\shortstack[c]{Neural}}}} 
& \multirow{2}{*}{NAF~\cite{luo2022learning}} & vanilla   &  0.64 & 0.71 & 0.021 & 10.08 & \multirow{2}{*}{\padd 5.51} & \multirow{2}{*}{22.04} & \multirow{2}{*}{11.98}  \\
& & \quad+ decay loss  &  0.64   & {\bf 0.53} & {\bf 0.017} & \padd 8.19 &  &  &  \\
\cdashlinelr{2-10}
& \multirow{2}{*}{INRAS~\cite{SuCS2022}} & vanilla  &  {\bf 0.36}   & 0.79 & 0.025 & \padd 8.01 & \multirow{2}{*}{{\bf \padd 1.33}}  & \multirow{2}{*}{{\bf 5.31}} & \multirow{2}{*}{3.36} \\
& & \quad + decay loss   &   0.39   &  0.57 & {\bf 0.017} & {\bf \padd 6.17} &  &  &  \\
\cdashlinelr{2-10}
& \multirow{2}{*}{NACF~\cite{liang2023neural}} & vanilla & 0.39     & 0.59 & {\bf 0.017} & \padd 6.62 & \padd 1.52 & 6.05 &  {\bf 3.17} \\
& & \quad + temporal &   0.39   & 0.59  &  0.018 & \padd 7.31 & \padd 1.75 & 7.00 & 3.41 \\
\cdashlinelr{2-10}
& AV-NeRF~\cite{liang2023av} & vanilla   &  0.39  & 0.73 & 0.021 & \padd 8.11 & 12.99 & 51.98 & 6.48\\
\bottomrule
\end{tabular}
}
\supparxiv{\vspace{-3.0mm}}{}
\end{table*}

\subsection{Sim2real for Few-Shot RIR Synthesis}

Real-world impulse responses can be expensive to acquire in large quantities, and obtaining a dense capture of such data for scenes can be particularly challenging.
For example, in comparison to a visual NeRF, there are comparatively fewer geometric constraints.
To address this limitation, we propose a two-stage training approach that leverages simulated audio data to enhance the synthesis of real-world audio with a limited amount of training samples.
Our method comprises two key stages: pretraining on dense synthetic data and fine-tuning on sparse real-world samples, as shown in \cref{fig:sim2real_method}.

\paragraph{Pretraining on dense synthetic data.}

In the first stage, we pretrain our audio neural field $\mathcal{F}$, \eg, INRAS, on the rich synthetic impulse responses generated from an acoustic simulator with diverse emitter and listener positions.
We use the room's geometry and acoustic properties~(reverberation) observed from limited real examples to create the simulator.
By exposing the model to a diverse range of simulated audio data, we enable it to learn general impulse response patterns and spatial information, which serve as a strong foundation for subsequent fine-tuning.

\paragraph{Fine-tuning on sparse real-world samples.}

We use sparse real-world audio samples for fine-tuning the neural field $\mathcal{F}$.
By fine-tuning it on real-world data, the model adapts to the specifics of real-world audio while retaining the knowledge gained from the simulator data.
By combining the strengths of the simulator and real-world data, our method achieves high-quality audio synthesis with sparse real-world data and strikes a balance between data collection cost and synthesis performance.

\section{Experiments}
\label{sec:exp}

We use our real-world dataset to evaluate audio and audio-visual acoustic field modeling methods. Also, we showcase our sim2real method that boosts them in the few-shot setting.

\begin{figure*}[t]
    \centering
    \upvspacefig
    \supparxiv{\vspace{-5.0mm}}{}
    \includegraphics[width=\linewidth]{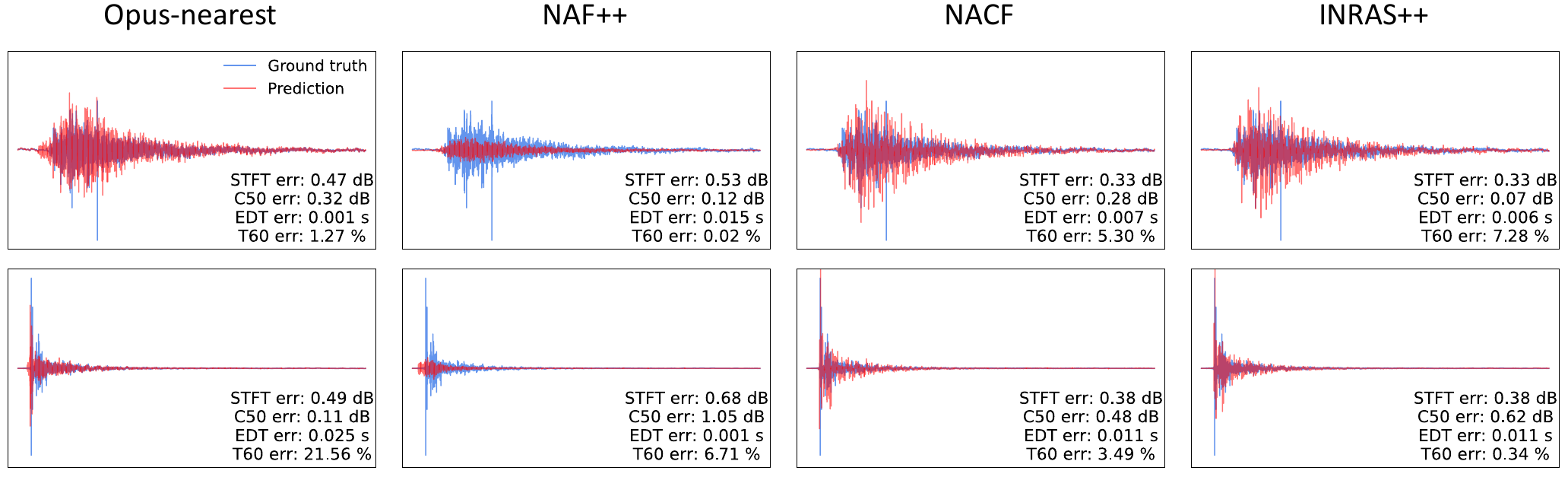}
    \caption{\textbf{Visualization of generated RIRs from different methods.}
        We visualize the ground-truth~(in blue) and predicted~(in red) impulse responses of several methods for qualitative comparison.
    }
    \label{fig:rir_comparison}
\end{figure*}

\subsection{Evaluation of 3D Neural Acoustic Fields}
\label{exp:benchmark}

We evaluate the methods on the 3D acoustic field modeling task using our full real-world dataset.

\paragraph{Models}

We consider both state-of-the-art neural field models and classical models.
To adapt to the 3D domain, we extend NAF~\cite{luo2022learning}, INRAS~\cite{SuCS2022}, NACF~\cite{liang2023neural}, AV-NeRF~\cite{liang2023av}, and their variants, introducing an extra dimension for neural acoustic field modeling, which was not feasible with other existing datasets.
Following \cite{luo2022learning, SuCS2022}, we also compare with traditional signal processing methods using linear and nearest-neighbor interpolation on the training data.
To improve the storage efficiency, we also apply audio encoding methods such as Advanced Audio Coding~(AAC) and Opus to compress audio with low bit rates; see \supparxiv{supp.}{\cref{supp:implement}} for details.

\paragraph{Metrics}

Following \citet{SuCS2022}, we use several metrics to assess the quality of the predicted impulse responses, including Clarity~($C_{50}$), Reverberation Time~($T_{60}$), and Early Decay Time~(EDT). $C_{50}$ quantifies the clarity of acoustics by measuring the ratio of initial sound energy to subsequent reflections within a room, with higher values indicating clearer acoustics.
$T_{60}$ reflects the overall sound decay within a room, while EDT focuses on the early portion of the sound decay curve.
We also evaluate STFT error,  the absolute error between the predicted and the ground-truth log-magnitude spectrograms~\cite{luo2022learning,defossez2018sing}.
Additionally, we measure the computational efficiency of each method by evaluating storage requirements for saving audio scenes and the inference time needed for rendering an impulse response.

\paragraph{Experimental setup}

For each scene, we use 80\% of the data for training and hold out 5\% and 15\% for validation and testing, respectively.
The impulse responses are resampled to 48\,kHz or 16\,kHz sampling rate and are cut to 0.32s for training and evaluation based on the average reverberate time of the room. 
For all experiments, we use the AdamW optimizer \cite{kingma2015adam,loshchilov2017decoupled} with a learning rate of $10^{-3}$, an exponential decay learning rate scheduler with a rate of 0.98, and a batch size of 128.
We train all the models on an NVIDIA A100 GPU for 200 epochs and evaluate the last epoch.
For NACF~\cite{liang2023neural} and AV-NeRF~\cite{liang2023av}, we use the visual NeRF model to render the corresponding RGB and depth images for novel views.
We test inference time on the same NVIDIA A100 GPU for all the methods to ensure fair comparison.

\begin{figure*}[t]
    \centering
    \upvspacefig
    \supparxiv{\vspace{-5.0mm}}{}
    \includegraphics[width=\linewidth]{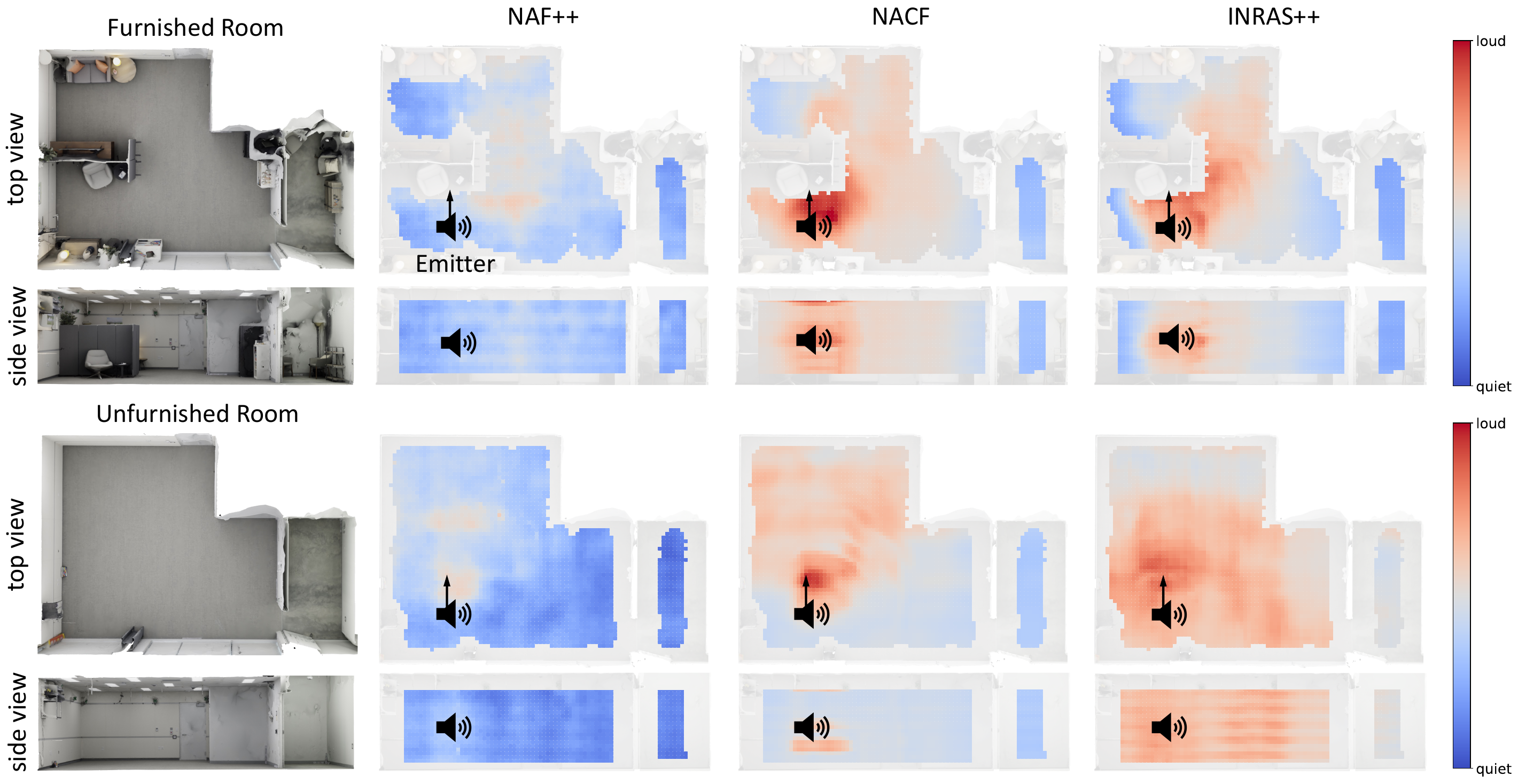}
    \caption{\textbf{Loudness map visualization.} Given an emitter position and its orientation, we visualize the intensity of predicted impulse responses over the spaces from the top view and side view, for the furnished and unfurnished room. Red means loud and blue means quiet. The arrow denotes the speaker's orientation.} 
    \label{fig:loudness_map}
    \downvspacefig
\end{figure*}

\paragraph{Results}
We show our quantitative results with 48K sampling rate RIRs in \cref{tab:benchmark_48k}.
We found that INRAS$++$, the version of INRAS with a decay loss, performs best on most metrics, has a lightweight architecture, and fast inference speed.
Opus audio encoding with nearest-neighbor interpolation is on par with several learning-based methods, suggesting the dense distribution of our captured RIRs throughout the scenes.
INRAS$++$ and NAF$++$ outperform their vanilla models by a large margin on the $C_{50}$, EDT, and $T_{60}$ metrics which indicates that the energy decay loss helps models learn the energy attenuation significantly better.
We show the qualitative result in \cref{fig:rir_comparison}.
We see that NACF and INRAS$++$ can generate impulse responses closer to ground truth while NAF$++$ fails even despite having good metric results.
We also visualize loudness maps in \cref{fig:loudness_map}, where we obtain 3D occupancy grids from our visual NeRF model with a resolution of 0.1\,m.
We can see that both INRAS$++$ and NACF learn a continuous acoustic field.
When comparing with the acoustic fields in furnished and unfurnished rooms, we can see the models have successfully captured the phenomenon of sound occlusion.
See \supparxiv{supp.}{\cref{supp:exp}} for more results.

\begin{figure}[!b]
    \centering
    \upvspacefig
    \supparxiv{\vspace{-0mm}}{}
    \includegraphics[width=\linewidth]{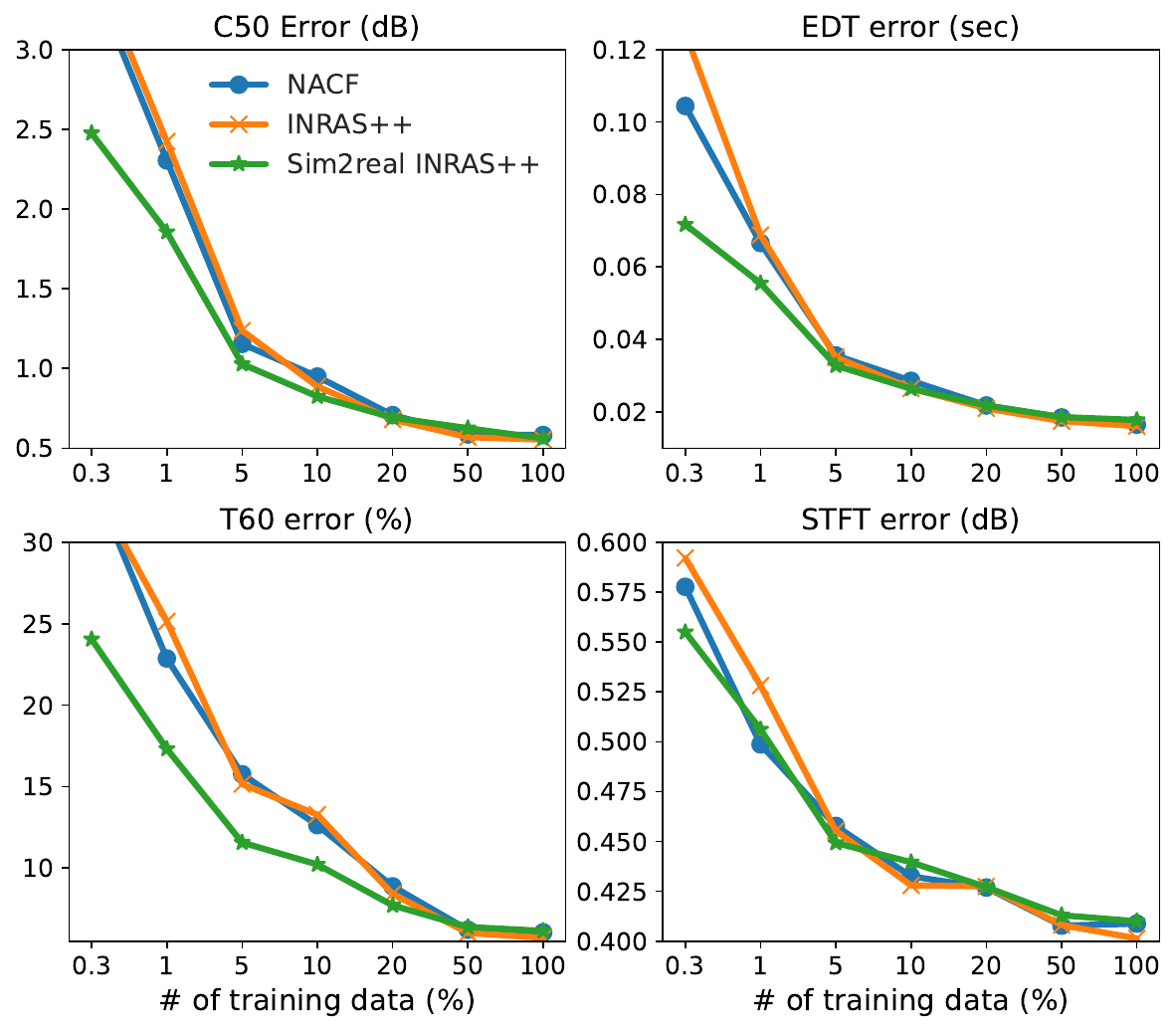}
    
    \caption{\textbf{Few-shot RIR synthesis results.} We evaluate the performances of models with different numbers of training data. The results are reported in the furnished room. Our Sim2Real method can improve the performance in cases of limited training data.
    } 
    \label{fig:sim2real_res}
    \downvspacefig
\end{figure}

\subsection{Evaluation of Few-Shot RIR Synthesis}

We next conduct experiments to explore how model performance varies with different training data scales.
Additionally, we benchmark our Sim2Real method against other baselines in challenging few-shot scenarios.

\paragraph{Experimental setup}

We trained each model on the furnished room using various training data scales, ranging from 0.3\% to 100\%, where the former comprises $\sim$100 samples and the latter has 31.3K samples.
To prevent overfitting, we reserved 10\% of the training samples for early stopping at each scale.
For our sim2real model, we created a geometric-based Pyroomacoustics~\citet{scheibler2018pyroomacoustics} shoebox acoustic simulator for pretraining, using the parameters of room bounding box and average $T_{60}$ calculated from real-world examples.
The training involved a dense pretraining stage on simulated data, followed by fine-tuning using real-world examples with a learning rate of $5 \!\times\! 10^{-4}$.

\paragraph{Results}

We present our results in \cref{fig:sim2real_res} and \cref{tab:fewshot}.
Our Sim2Real model demonstrates substantial performance improvements in few-shot setups, specifically with 0.3\%, 1\%, and 5\% of the total training samples~(approximately 100, 300, and 1500 samples, respectively).
As the number of training samples increases, the advantages of using simulated data for training become smaller.
While our sim2real model lags behind the model trained with the complete dense dataset, which exhibits a 55\% improvement over our model trained with 5\% of the data, it's worth noting that we only use a basic shoebox simulator.
We believe this performance gap will be further narrowed down if we can apply more advanced simulators, such as \citet{ChenJSGAIRG2020,chen2022soundspaces}.

\begin{table}[t!]
\caption{\textbf{Few-shot experiments with 1\% training data.}}
\label{tab:fewshot}

\centering
\resizebox{1\columnwidth}{!}{%
    \begin{tabular}{lcccc}
    \toprule
         \multirow{2}{*}{Method} &  STFT Err & $C_{50}$ Err & EDT Err & $T_{60}$ Err  \\
 & (dB)~$\downarrow$ & (dB)~$\downarrow$ & (sec)~$\downarrow$ & (\%)~$\downarrow$ \\
    \midrule
        Simulator~\cite{scheibler2018pyroomacoustics} & 1.47 &  5.14  &  0.170 & 44.00 \\
        NAF$++$ &  0.69 & 2.38 &  0.072 & 21.77 \\
        INRAS$++$ &  0.53  & 2.42  &  0.098 & 25.15 \\
        NACF &  {\bf 0.50}  & 2.30  &  0.067 & 22.87 \\
        sim2real INRAS$++$ & 0.51  & {\bf 1.86 }	& {\bf 0.056} &	{\bf 17.31}  \\
    \bottomrule
    \end{tabular}%
}
\downvspacefig
\end{table}

\subsection{Ablation Study}
\begin{table}[!t]
\caption{\textbf{Ablation experiments on the bounce point sampling strategy.} We perform the experiments on the furnished room.}
\label{tab:bnc_sampling}

\centering
\resizebox{1\columnwidth}{!}{%
    \begin{tabular}{lcccccc}
    \toprule
       \multirow{2}{*}{Method} & \multirow{2}{*}{Modality} & Bnc point & STFT & $C_{50}$ & EDT & $T_{60}$   \\
  &  & strategy & (dB)~$\downarrow$ & (dB)~$\downarrow$ & (sec)~$\downarrow$ & (\%)~$\downarrow$ \\
    \midrule
        INRAS$++$ & $\mathcal{A}$ & 2D &  0.42  &  0.57   & 0.017   & 6.21 \\
        NACF  & $\mathcal{A}$ \& $\mathcal{V}$  & 2D &  {\bf 0.40}  & 0.58  & 0.017   & 6.04 \\
        INRAS$++$ & $\mathcal{A}$ & 3D &  0.41  & {\bf 0.53}   & {\bf 0.016}   & {\bf 5.84} \\
        
    \bottomrule
    \end{tabular}%
}
\supparxiv{\vspace{-3.5mm}}{}
\end{table}
\paragraph{Bounce point sampling}

We investigate the impact of our bounce point sampling strategy on model performance.
To do this, we compare our 3D bounce point sampling method with the original 2D sampling method at a fixed height.
As shown in \cref{tab:bnc_sampling}, our 3D bounce point sampling enhances the model's performance.
Additionally, INRAS$++$ with 2D bounce points shows a comparable performance against the audio-visual model NACF, suggesting that the audio modality alone suffices for the current setup.

\begin{table}
\supparxiv{\vspace{1mm}}{}
\caption{\textbf{Ablation experiments on speaker orientation.} }
\label{tab:speak_orientation}

\centering
\resizebox{1.0\columnwidth}{!}{%
    \begin{tabular}{lcccc}
    \toprule
         \multirow{2}{*}{Method} &  STFT Err & $C_{50}$ Err & EDT Err & $T_{60}$ Err  \\
 & (dB)~$\downarrow$ & (dB)~$\downarrow$ & (sec)~$\downarrow$ & (\%)~$\downarrow$ \\
    \midrule
        INRAS$++$ w/o ori. & 0.42  & 0.64  & 0.018  & 6.57  \\
        INRAS$++$  &  {\bf 0.40}  & {\bf 0.55}	& {\bf 0.016} &	{\bf 5.59}  \\
    \bottomrule
    \end{tabular}%
}
\supparxiv{\vspace{-5mm}}{}
\end{table}

\paragraph{Speaker orientation.} 
We use a directional speaker during our data capture, exhibiting directivity patterns that affect the acoustic experience of receivers
We explore how neural models use orientation information by removing speaker orientation embeddings from the inputs.
As demonstrated in \cref{tab:speak_orientation}, the quality of the generated RIRs significantly improves when orientation information is included.
Please see \supparxiv{supp.}{\cref{supp:dataset}} for RIR visualization for different orientations.

\paragraph{Energy decay loss}

We study how model performance varies with the weights of the energy decay loss.
We conduct experiments for INRAS$++$ on the furnished room and set $\lambda$ to $\{0.1, 0.2, 0.3, 0.5\}$.
We present our results in \cref{tab:decay_loss}.
It shows that increasing the weights of decay loss improves metrics like $C_{50}$, EDT, and $T_{60}$ errors, though it comes with a tradeoff in the STFT error metric. For our primary experiments, we choose $\lambda = 2.0$ for balanced performance.

\section{Conclusion}

This paper introduces \ourdata, a multimodal real-world acoustic room dataset collected for facilitating research on novel-view acoustic synthesis and neural acoustic field modeling techniques. \ourdata includes dense 3D room impulse response captures of a large space, both with and without furniture. It also include visual data captured from multiple viewpoints and precise tracking of sound sources and receivers in the room. We systematically evaluated existing techniques for audio and audio-visual novel-view acoustic synthesis using this real-world data. We provided insights into the performance of individual models and proposed new improvements. Furthermore, we conducted experiments to investigate the impact of incorporating visual data (\ie, images and depth) into neural acoustic field models.
This dataset fills a gap in existing research by providing real-world data for evaluating and benchmarking novel-view acoustic synthesis models and impulse response generation techniques. In the future, we plan to expand the dataset to more room configurations.

\begin{table}[t!]
\caption{\textbf{Ablation experiments on the energy decay loss.} }
\label{tab:decay_loss}

\centering
\resizebox{0.9\columnwidth}{!}{%
    \begin{tabular}{lccccc}
    \toprule
         & \multirow{2}{*}{$\lambda$} & STFT Err & $C_{50}$ Err & EDT Err & $T_{60}$ Err  \\
  &  & (dB)~$\downarrow$ & (dB)~$\downarrow$ & (sec)~$\downarrow$ & (\%)~$\downarrow$ \\
    \midrule
        \multirow{4}{*}{INRAS$++$} & 1.0 &  {\bf 0.39}  &  0.58   & 0.017   & 5.98 \\
         & 2.0 &  0.40  & 0.55   &  0.016 & 5.59 \\
          & 3.0 &  0.41  & { \bf 0.49}   &  0.016   &  5.48 \\
          & 5.0 &  0.43  & { \bf 0.49}   & {\bf 0.015}   & {\bf 5.34}  \\
    \bottomrule
    \end{tabular}%
}
\supparxiv{\vspace{-5mm}}{}
\end{table}

\paragraph{Limitations and Broader Impacts}
Collecting real-world room impulse data is expensive and time-consuming, which makes scaling up data collection for multiple rooms or scenes challenging. Our dataset only have RIRs data from a single physical room, although with two different configurations. Thus, its utility is limited for research aiming to generalize across different rooms and scenes. Using RIR data can produce audio recordings that mimic real recordings from a specific room. However, this capability can lead to the creation of deceptive and misleading media.

\paragraph{Acknowledgements}
We would like to thank Jake Sandakly, Steven Krenn, and Todd Keebler for their assistance with data collection, as well as Linning Xu and Vasu Agrawal for their support with VR-NeRF rendering. Our gratitude also extends to Kun Sun, Mingfei Chen, Susan Liang, and Chao Huang for their insightful discussions.

{
    \small
    \bibliographystyle{ieeenat_fullname}
    \bibliography{main,Audiovisual-CR}
}

\clearpage

\supparxiv{
\setcounter{page}{1}
\maketitlesupplementary
}{}

\appendix

\renewcommand{\thesection}{A.\arabic{section}}
\setcounter{section}{0}

\section{Additional Experimental Results}
\label{supp:exp}

\paragraph{Benchmark on 16\,KHz impulse responses}

We also evaluate each method on our benchmark with impulse responses of 16\,kHz sampling rate.
We show the results in \cref{tab:benchmark_16k}.
We can see that INRAS$++$ performs best overall, which matches with the conclusion in \cref{exp:benchmark}.

\paragraph{More qualitative results}

\begin{figure*}[t]
    \centering
    \includegraphics[width=\linewidth]{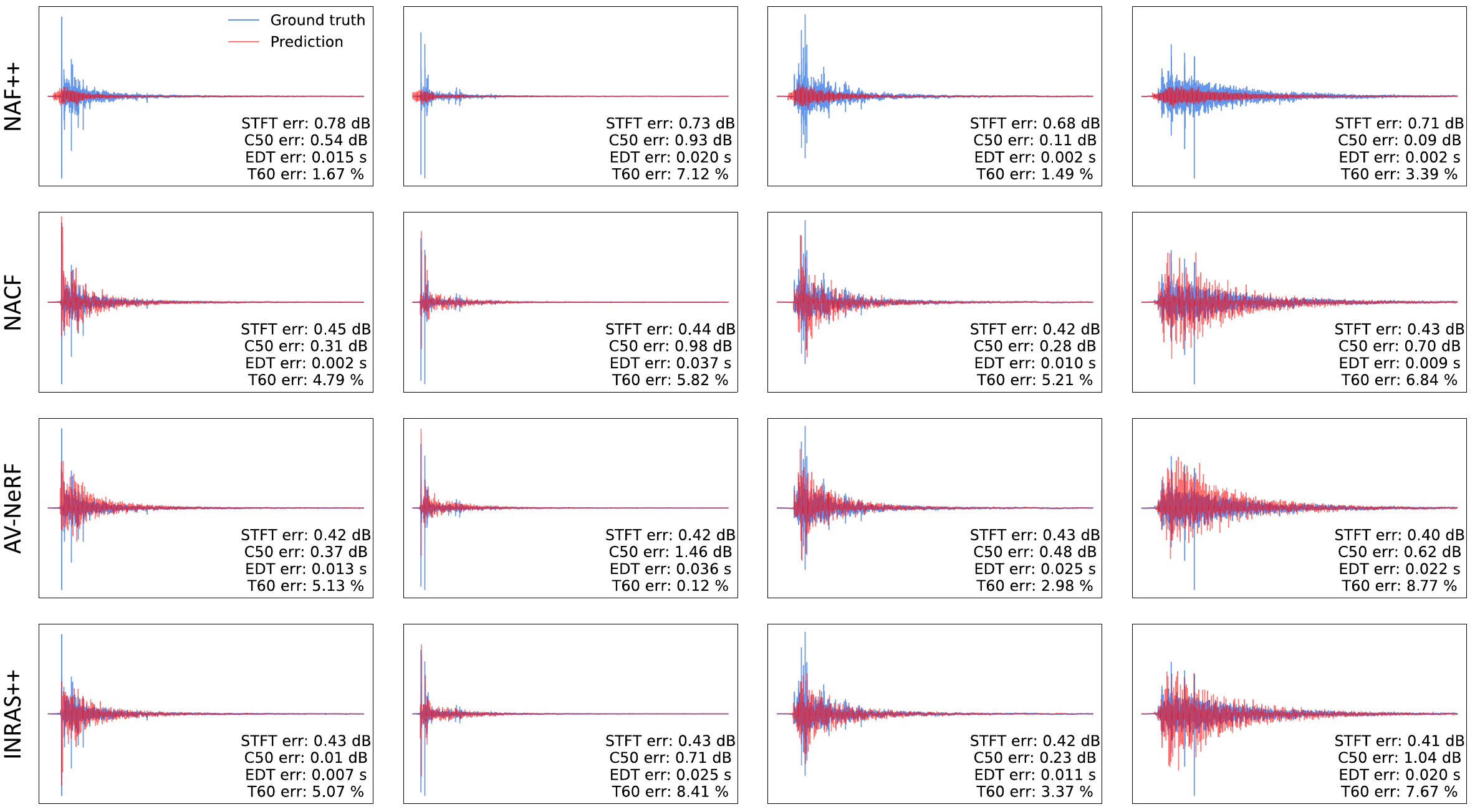}
    \caption{\textbf{Visualization of generated RIRs.} We visualize the ground truth~(in blue) and predicted~(in red) impulse responses of several methods for qualitative comparison. } 
    \label{fig:rir_comparison_supp}
\end{figure*}

\begin{figure*}[t]
    \centering
    \includegraphics[width=\linewidth]{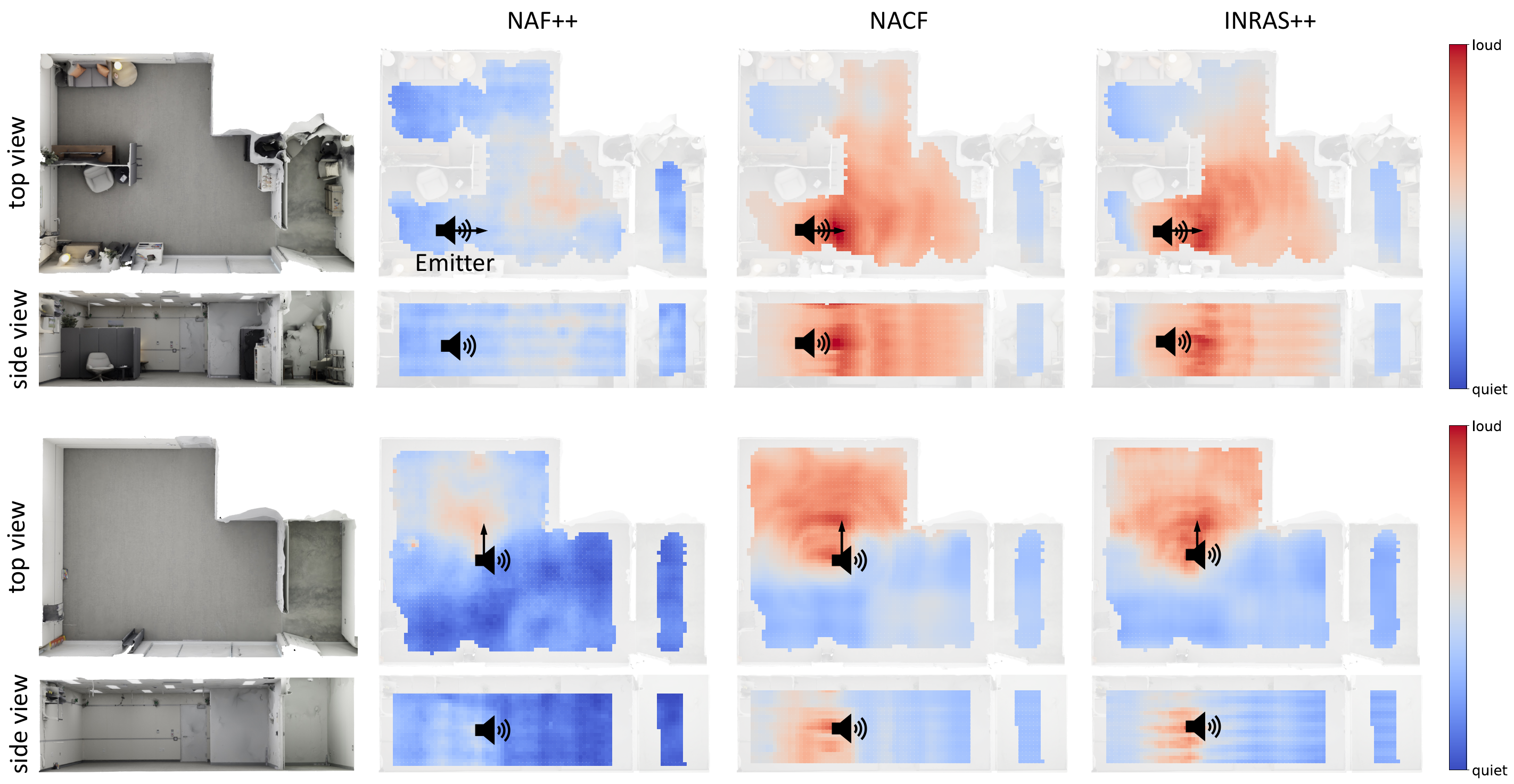}
    \caption{\textbf{Loudness map visualization.} We visualize the intensity of predicted impulse responses over the spaces from the top view and side view given an emitter position and its orientation. Red means loud and blue means quiet. } 
    \label{fig:loudness_map_supp}
\end{figure*}

We provide more predicted RIR visualization for qualitative comparison in \cref{fig:rir_comparison_supp}.
We also provide more loudness map visualization on the different scenes for qualitative comparison in \cref{fig:loudness_map_supp}.

\paragraph{Empty versus furnished room}

\begin{figure*}[t]
    \centering
    \includegraphics[width=\linewidth]{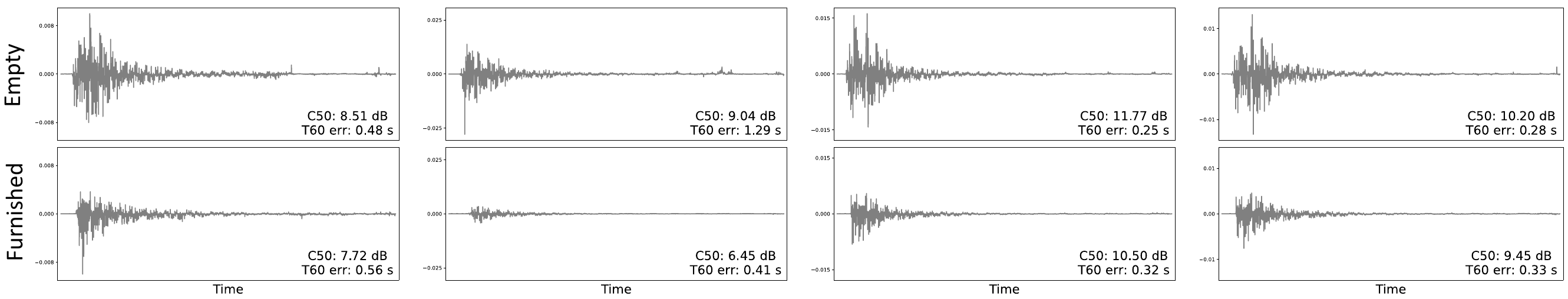}
    \caption{\textbf{Visualization comparison of generated RIRs from different scenes.} We present visualizations of four pairs of generated impulse responses, each sharing the same emitter-receiver position in both the empty room and the furnished room. These visualizations highlight the variations in the acoustic fields between the two distinct scenes.
    }
    \label{fig:rir_empty2furn}
\end{figure*}

One advantage of our dataset is that it contains a scene in two conditions --- empty and furnished, which allows studying the difference in acoustic fields introduced by furniture.
Due to a lack of ground-truth comparison, we visualize the generated impulse responses from INRAS++ trained on each scene individually as an approximation of the acoustic field.
We show our results in \cref{fig:rir_empty2furn}, where we can see that generated impulse responses with different acoustic properties.

\begin{table*}[!t]
\caption{\textbf{Benchmark with 16\,kHz sampling rate.}  }
\label{tab:benchmark_16k}

\centering
\newcommand\padd{\phantom{0}}

\resizebox{\textwidth}{!}{
\begin{tabular}{cll@{}cccccrr}

\toprule
 & \multirow{2}{*}{Method} & \multirow{2}{*}{Variation} & STFT error   & $C_{50}$ error & EDT error & $T_{60}$ error & Parameters &  Storage & Speed \\
 & & & (dB)~$\downarrow$  & (dB)~$\downarrow$ & (sec)~$\downarrow$ & (\%)~$\downarrow$ & (Million)~$\downarrow$ & (MB)~$\downarrow$ & (ms)~$\downarrow$ \\
\midrule

\parbox[t]{2mm}{\multirow{6}{*}{\rotatebox[origin=c]{90}{\shortstack[c]{Classical}}}} 
&  \multirow{3}{*}{Linear} & AAC &  1.14 & 1.09  & 0.040  &  8.79 & \multirow{3}{*}{--} & 680.45 & \multirow{3}{*}{--} \\
& & Opus &  1.06  & 0.80 & 0.032 & 7.48 &   & 680.45  &  \\
& & original &  1.02  & 0.82   & 0.032  &  6.82 &  &  3,172.77  &  \\

\cdashlinelr{2-10}

&  \multirow{3}{*}{Nearest} & AAC &   0.72  & 0.83 & 0.027 & 8.08 & \multirow{3}{*}{--} & 680.45 & \multirow{3}{*}{--} \\
& & Opus &  0.58  & 0.61 & 0.020 & 6.96 &  & 680.45  &  \\
& & original & 0.48  & 0.71   & 0.020  &  7.68 &  &  3,172.77  &  \\

\midrule
\parbox[t]{2mm}{\multirow{8}{*}{\rotatebox[origin=c]{90}{\shortstack[c]{Neural}}}} 
& \multirow{2}{*}{NAF~\cite{luo2022learning}} & vanilla  &  0.77   &  0.69 & 0.025 & 8.15 & \multirow{2}{*}{\padd  5.51} & \multirow{2}{*}{22.04} & \multirow{2}{*}{5.57}  \\
& & \quad+ decay loss &  0.77  & 0.63 & 0.023 & 7.43 &  &  &  \\
\cdashlinelr{2-10}
& \multirow{2}{*}{INRAS~\cite{SuCS2022}} & vanilla &  {\bf 0.44}  & 0.65 & 0.024 & 6.15 & \multirow{2}{*}{{\bf \padd  1.33}}  & \multirow{2}{*}{{\bf 5.31}} & \multirow{2}{*}{{\bf 2.10}} \\
& & \quad + decay loss &   0.45  & {\bf 0.54} & {\bf 0.019} & {\bf 5.34} &  &  &  \\
\cdashlinelr{2-10}
& \multirow{2}{*}{NACF~\cite{liang2023neural}} & vanilla  & 0.45   & 0.58 &  0.020 & 5.47 & \padd  1.52 & 6.05 &  2.39 \\
& & \quad + temporal &  0.48  & 0.60 & 0.022 & 6.59  & \padd  1.75 & 7.00 &  2.78 \\
\cdashlinelr{2-10}
& AV-NeRF~\cite{liang2023av} & vanilla  &  0.46   & 0.58 & 0.021 & 6.12 & 12.99 & 51.98 & 5.80\\
\bottomrule
\end{tabular}
}
\end{table*}

\begin{figure}[!hb]
    \centering
    \includegraphics[width=\linewidth]{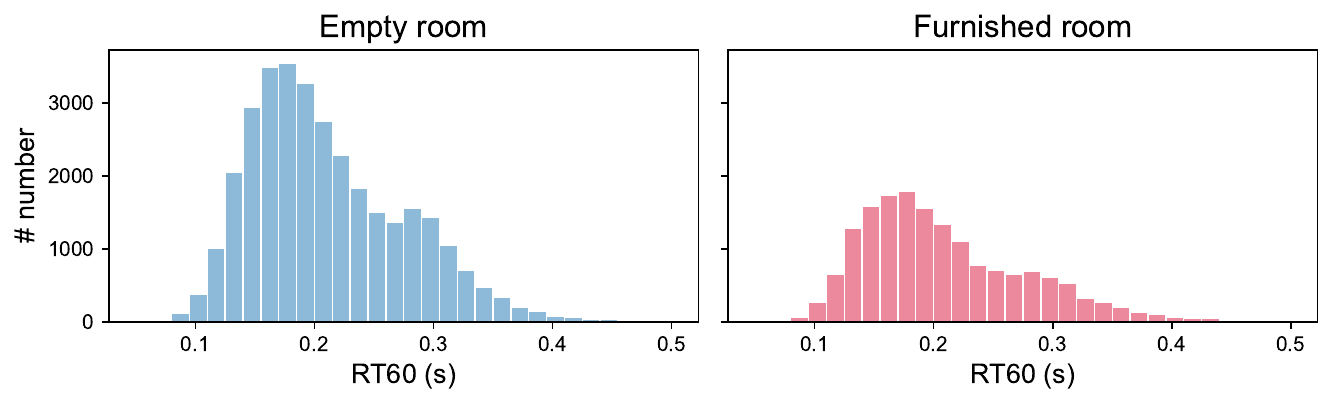}
    \caption{\textbf{RT60 distribution.}
    } 
    \label{fig:rt60}
\end{figure}


\section{Implementation Details}
\label{supp:implement}

In this section, we will demonstrate the implementation of each baseline in detail.

\paragraph{AAC and Opus}

We convert the raw waveform~({\tt .wav}) into AAC~({\tt .m4a}) and Opus ({\tt .opus}) encoding and reverse the compression using FFmpeg commands as shown below: 

\begin{lstlisting}[language=Python, caption=FFmpeg commands for audio compression]
# AAC compression
encode_command = f"ffmpeg -i audio.wav -t {audio_length} -c:a aac -b:a 24k temp.m4a"
decode_command = f"ffmpeg -i temp.m4a -c:a pcm_f32le -ar {sampling_rate} audio_aac.wav"

# Opus compression
encode_command = f"ffmpeg -i audio.wav -t {audio_length} -c:a opus -strict -2 -b:a 24k temp.opus"
decode_command = f"ffmpeg -i temp.opus -c:a pcm_f32le -ar {sampling_rate} audio_opus.wav"
\end{lstlisting}

We cut the audio to be the same length~(0.32s) and corresponding sampling rate (16K or 48K) for fair evaluation comparison.

Note that we use a different Opus encoder which can achieve better compression performance than NAF used~\cite{luo2022learning}.
Due to the heavy computation of constructing a high-dimensional interpolation engine, we modify the baseline algorithm by first matching the nearest neighbor of the emitter in the training distribution and then performing the nearest neighbor or linear interpolation to generate impulse responses for given listener positions.

\paragraph{NAF}

We follow the official implementation of NAF~\footnote{\href{https://github.com/aluo-x/Learning_Neural_Acoustic_Fields/}{https://github.com/aluo-x/Learning\_Neural\_Acoustic\_Fields/}}, and create 3D grid features based on the bounding boxes of scenes.
For experiments with 16\,kHz sampling rate, we use an STFT with an FFT size of 512, a window length of 256, and a hop length of 128.
For 48\,kHz sampling rate, we use an STFT with an FFT size of 1024, a window length of 512, and a hop length of 256.
We perform inverse STFT on the predicted magnitude of the RIR spectrogram with random spectrogram phase to obtain time-domain RIR. We set $\lambda=1.0$ for the weight of energy decay loss when training NAF$++$.

\paragraph{INRAS}

We follow the implementation of INRAS provided by the authors in their supplementary material\footnote{\href{https://openreview.net/forum?id=7KBzV5IL7W}{https://openreview.net/forum?id=7KBzV5IL7W}}, and add an extra dimension for the emitter, listener, and bounce point position.
We changed the original bounce point sampling method, which only sampled points with a specific height.
Instead, we apply Poisson sampling on the scene meshes to obtain 256 bounce points in 3D to represent scene geometry in a better way.
To optimize multi-resolution STFT loss, we set FFT size as $\{128, 512, 1024, 2048\}$, window length as $\{80, 240, 600, 1200\}$, and hop length as $\{16, 50, 120, 240\}$.
We set $\lambda=2.0$ for the weight of the energy decay loss.

\paragraph{NACF}

We use the same architecture as INRAS for NACF.
We keep the original bounce point sampling strategy in the paper and render visual context using VR-NeRF~\cite{XuALGBKBPKBLZR2023}.
We render 256$\times$256\,pixel RGB, and depth images with a field of view of 90°.
We use the surface normal of each bounce point to determine the look-at view of the virtual camera.
Following the original paper, RGB and depth images are down-sampled to $16 \times 16$ and are encoded with an MLP as visual contexts.
We set $\lambda=2.0$ for the weight of energy decay loss.
We optimize the multi-resolution STFT loss with the same hyperparameters as INRAS. 

\paragraph{AV-NeRF}

Because we have a different setup from AV-NeRF~\cite{liang2023av} where we have omnidirectional microphones instead of orientated binaural receivers, we adopt their method with several changes.
We use VR-NeRF \cite{XuALGBKBPKBLZR2023} to render 4 perspective views of 256$\times$256 RGB and depth maps with a field of view of 90° for each receiver's position, and encode them with frozen ResNet18~\cite{he2016deep} trained on ImageNet~\cite{deng2009imagenet}.
We removed the relative angle because it does not fit our setup.
We set $\lambda=2.0$ for the weight of energy decay loss.

\section{Dataset}
\label{supp:dataset}

\paragraph{Impulse response data processing}
We followed the sine-sweep deconvolution process as described by \citet{farina2000simultaneous} to extract the impulse response from the signals recorded by the microphones.
For each extracted impulse response, we saved the 3D location of the receiver, as well as the 3D location and orientation of the sound source. The length of the impulse response is 4 seconds and all audio data was recorded at a sampling rate of 48\,kHz and stored at a resolution of 32 bits. We show the RT60 distribution of our collected RIRs in \cref{fig:rt60}

\paragraph{Visual rendering}

\begin{figure*}[t]
    \centering
    \includegraphics[width=\linewidth]{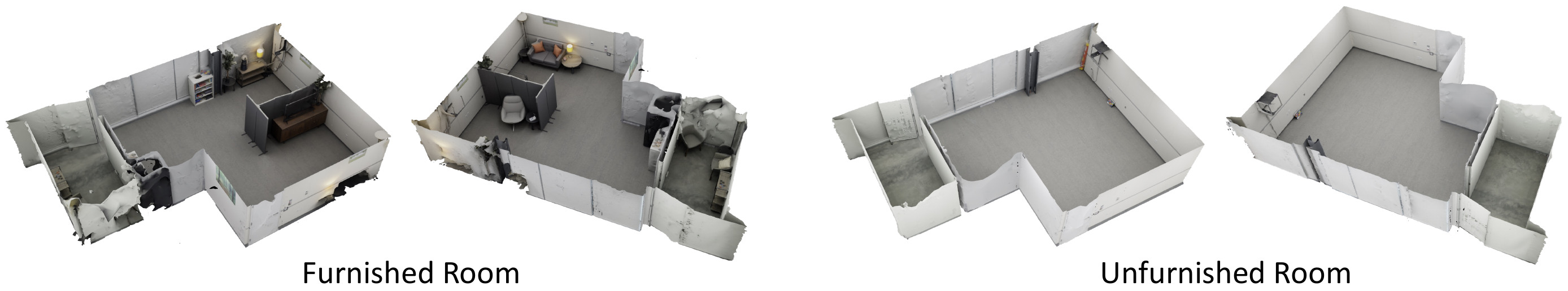}
    \caption{\textbf{Scene overview of \ourdata.} } 
    \label{fig:scene_overview}
\end{figure*}

We provide renderings of room meshes as a simple overview in \cref{fig:scene_overview}.

\paragraph{Speaker orientation}

\begin{figure*}[t]
    \centering
    \includegraphics[width=\linewidth]{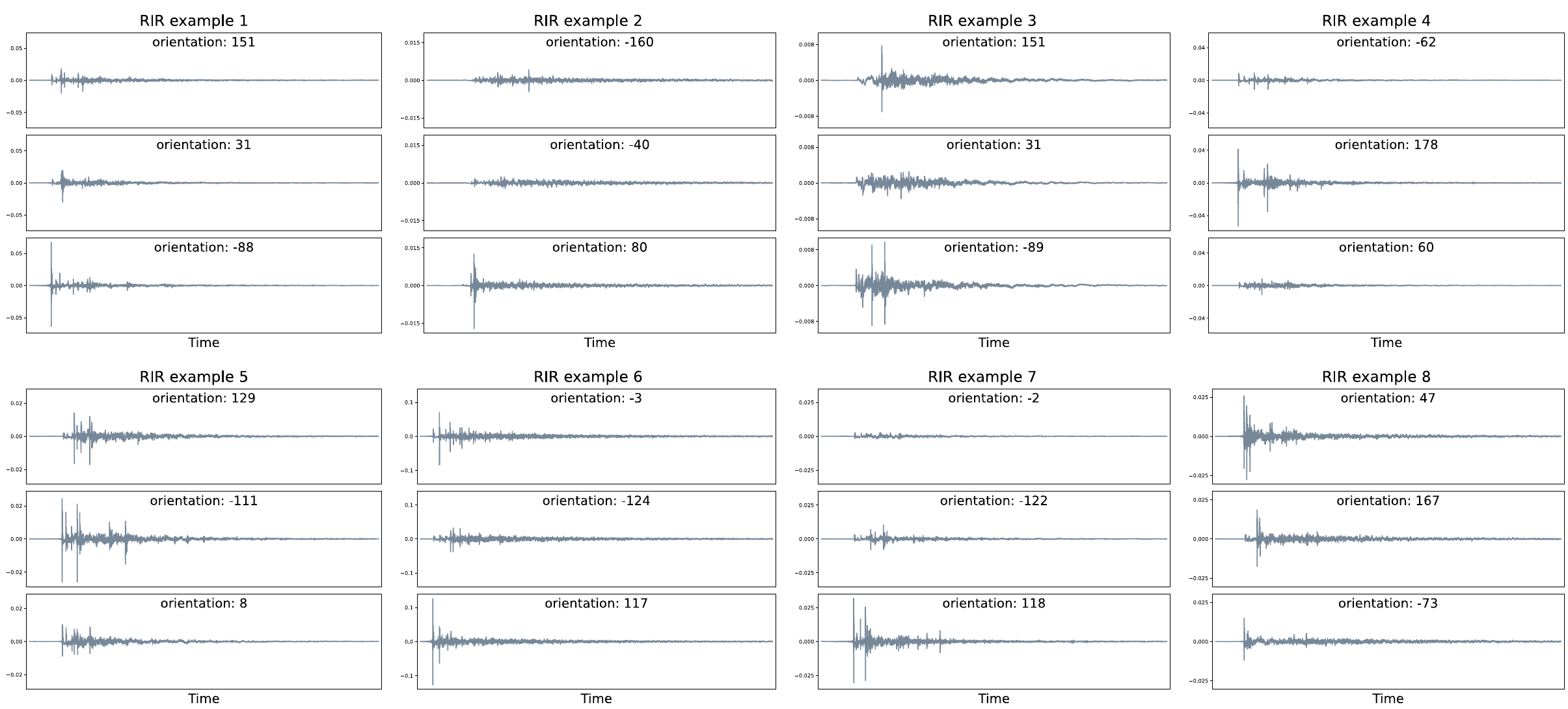}
    \caption{
        \textbf{Visualization of ground-truth RIRs with different orientations.}
    }
    \label{fig:orientation_rir}
\end{figure*}

In \cref{fig:orientation_rir}, we provide visualizations of impulse response pairs from our captured dataset.
These pairs share the same emitter-listener position but differ in emitter orientations.
The orientations of directional speakers impact the resulting impulse responses.

\end{document}